\newcommand\Dioxane{2Dioxane$\cdot$2H$_2$O$\cdot$CuCl$_2$}
\documentclass[twocolumn,prb,showpacs,preprintnumbers,amsmath,amssymb,superscriptaddress]{revtex4}

\usepackage{graphicx}
\usepackage{dcolumn}
\usepackage{bm}
\usepackage{epsfig}
\begin{document}

\title{Synthesis and structural characterization of \Dioxane: metal-organic compound with Heisenberg antiferromagnetic S=1/2 chains}

\author{Tao Hong}
\email[Electronic address: ] {hongt@ornl.gov}
\author{R. Custelcean}
\author{B. C. Sales}
\affiliation{Oak Ridge National Laboratory, Oak Ridge, Tennessee 37831-6393, USA.}
\author{B. Roessli}
\affiliation{Laboratory for Neutron Scattering ETHZ and Paul Scherrer Institut, CH-5232 Villigen PSI, Switzerland}
\author{D. K. Singh}
\affiliation{NIST Center for Neutron Research, National Institute of Standards and Technology, Gaithersburg, MD 20899, USA.}
\affiliation{Department of Materials Science and Engineering, University of Maryland, College Park,
Maryland 20742, USA.}
\author{A. Zheludev}
\affiliation{Oak Ridge National Laboratory, Oak Ridge, Tennessee 37831-6393, USA.}
\affiliation{Laboratory for Neutron Scattering ETHZ and Paul Scherrer Institut, CH-5232 Villigen PSI, Switzerland}

\date{\today}

\begin{abstract}
A novel organometallic compound \Dioxane~has been synthesized and structurally characterized by X-ray crystallography. Magnetic susceptibility and zero-field inelastic neutron scattering have also been used to study its magnetic properties. It turns out that this material is a weakly coupled one-dimensional S=1/2 Heisenberg antiferromagnetic chain system with chain direction along the crystallographic \textbf{c} axis and the nearest-neighbor intra-chain exchange constant \emph{J}=0.85(4) meV. The next-nearest-neighbor inter-chain exchange constant $J'$ is also estimated to be 0.05 meV. The observed magnetic excitation spectrum from inelastic neutron scattering is in excellent agreement with numerical calculations based on the M$\ddot{u}$ller ansatz.
\end{abstract}


\maketitle

Progress in the field of quantum magnetism is driven by advances in synthesis of prototypical materials. One of them is the S=1/2 trimer chain system.\cite{Hida63:94,Okamoto98:96,Ishii69:00} One realization of such model was made by Ajiro \emph{et al.} in the material 2Dioxane$\cdot$3CuCl$_2$.\cite{Ajiro63:94} After following the published sample preparation procedure as described in the Ref.[5], we found this compound to be unstable under ambient conditions. Instead, a novel compound, \Dioxane~(CuDCl) was obtained. Here, we describe the crystal structure and magnetic property characterization of CuDCl, which turns out to be a weakly coupled quasi one-dimensional (1D) Heisenberg antiferromagnetic (HAF) S=1/2 chain. The physical properties of HAF S=1/2 chain system have been thoroughly studied in a number of materials, such as KCuF$_3$\cite{Nagler44:91,Tennant70:93,Tennant52:95}, $\rm Cu(C_6D_5COO)_2\cdot3D_2O$ (Copper Benzoate)\cite{Dender53:96,Dender79:97}, BaCu$_2$Si$_2$O$_7$\cite{Tsukada60:99,Zheludev85:00,Kenzelmann64:01,Zheludev65:01,Zheludev89:02,Zheludev67:03}, $\rm Cu(C_4H_4N_2)(NO_3)_2$ (CuPzN)\cite{Hammar59:99,Stone91:03}. However, unlike these compounds, the Cl concentration in CuDCl can be arbitrarily tuned by Br substitution. As a result, the strength of superexchange coupling \emph{J} changes dramatically with different Br concentrations \emph{x}.\cite{Hong09} The new compound opens a new research field based on the HAF S=1/2 chain system with bond randomness.

We have determined the crystal structure of CuDCl ($\rm C_8H_{20}O_6CuCl_2$) by X-ray diffraction at \emph{T}=173 K. The compound crystallizes in the monoclinic space group C2/c with lattice constants $a$=17.4298(9) $\AA$, $b$=7.4770(4) $\AA$, $c$=11.8230(6) $\AA$, and $\beta$=119.4210(10)$^\circ$.

\begin{figure}
 \includegraphics[width=1.0\columnwidth]{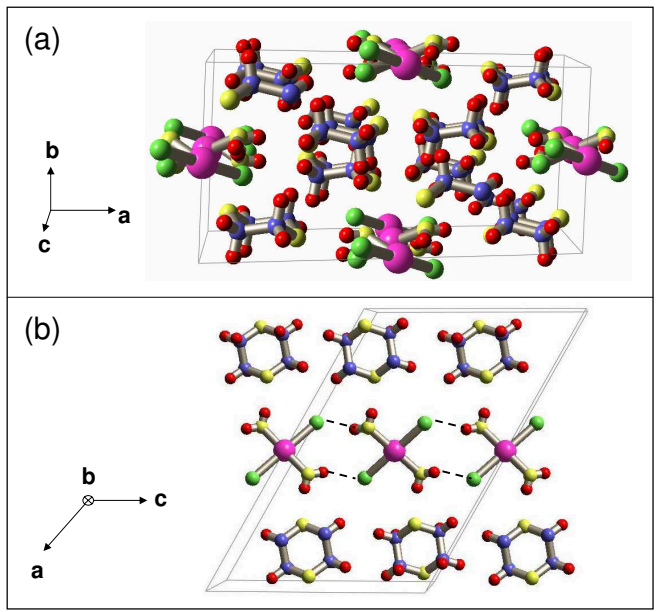}
 \caption{(color online) Crystal structure of CuDCl (a) View along the \textbf{c} axis showing the well separated bi-layers structure. (b) View along the \textbf{b} axis showing how Cu$^{2+}$ ions are linked into a 1D chain. The chain axis is along the \textbf{c} direction. Dashed lines show the super-exchange Cu-Cl-H-O-Cu pathway. The grey outline is one unit cell. Color coding is as follows: Pink:Cu, Green:Cl, Blue:C, Red: H, and Yellow:O.}
 \label{struc}
\end{figure}

The packing structure can be described as composed of CuCl$_2$$\cdot$2H$_2$O bi-layers in the \textbf{a-c} plane as shown in Fig.~\ref{struc}(a). They are well separated by the 1,4-Dioxane organic molecules. Hence the possible mediated inter-layer interactions existing between molecular units displaced by $(\bf{a}\pm\bf{b})/2$ should be very weak. Fig.~\ref{struc}(b) shows the structure in a single \textbf{a-c} plane. The Cu$^{2+}$ ions form chains along the crystallographic \textbf{c} axis and the distance between nearest neighbor Cu$^{2+}$ ions is c/2. The chains are separated from each other by the 1,4-Dioxane molecules. The Cu$^{2+}$ ions have a 4 coordination with their neighboring chlorine and oxygen atoms. The four short bonds including two Cu-Cl (2.285 $\AA$) and two Cu-O bonds (1.959 $\AA$) are coplanar, thus defining the $\hat{\bf x}$-$\hat{\bf y}$ plane for the $3d_{x^2-y^2}$ copper orbital. Thus the exchange constant coupling the nearest Cu$^{2+}$ ions through the super-exchange pathway Cu-Cl-H-O-Cu (dash lines) should be dominant.

Crystals of CuDCl were grown by slow evaporation of aqueous solutions containing anhydrous Cu(II) chloride and excess of 1,4-Dioxane and methanol. Blue single crystals with a shape of planar irregular hexagon were harvested after a few weeks.

 X-ray diffraction data were collected on a tiny single crystal sample using Bruker SMART Apex CCD diffractometer (Mok$\alpha$ radiation, $\lambda$=0.71073$\AA$) using $\phi$ and $\omega$ scans. Totally 6122 reflections were collected. The structure was solved via direct methods using SHELX97 and refined via least-squares using SHELX97.

DC magnetic susceptibility measurement was performed on a single crystal sample (\emph{m}=0.03 g) in the temperature range of 1.8 K$<T<$300 K using a superconducting quantum interference device (SQUID) magnetometer in a DC field of 0.1 Tesla.

Inelastic neutron scattering (INS) measurements on cold neutron triple axis spectrometer TASP,\cite{Semadeni297:01} at the SINQ spallation source were performed using one deuterated single crystal with a mass of \emph{m}=0.6 g. In this Setup, the spectrometer was operated in fully open horizontal collimation mode with final energy fixed at 5 meV. Experiment in Setup 2 was performed on SPINS at the NIST Center for Neutron Research (NCNR). The sample consists of two co-aligned single crystals with total mass 0.5 g and a mosaic of 1$^\circ$. Open horizontal beam divergences through the triple axis spectrometer were given by ($^{58}$Ni-guide)-80$^\prime$-300$^\prime$-240$^\prime$ collimations. A horizontally-focusing pyrolytic graphite (PG 002) analyzer set to reflect neutrons with final energy $E_f=3.7$ meV was used. A liquid-nitrogen-cooled Be (Setup 1) or BeO (Setup 2) filter was placed after the sample to eliminate higher-order beam contamination. In both Setups, sample was oriented with its reciprocal $(h,0,l)$ plane in the horizontal plane and the temperature was controlled by the conventional liquid helium cryostat. Wave-vector transfer is indexed as ${\bf q}=h{\bf a}^\ast+l{\bf c}^\ast$.

\begin{figure}
 \includegraphics[width=7.5cm,bbllx=90,bblly=375,bburx=545,
  bbury=690,angle=0,clip=]{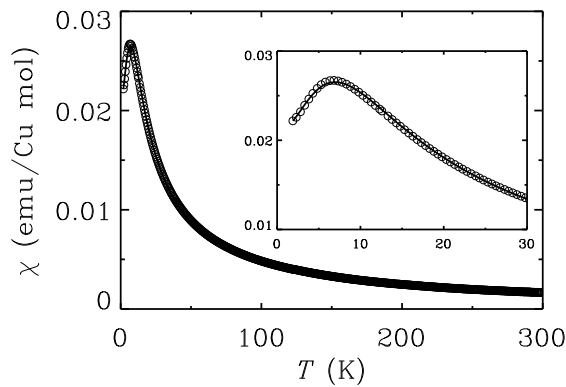}
 \caption{Magnetic susceptibility as a function of temperature in a single crystal sample of CuDCl. Inset: enlarged low-temperature part of magnetic susceptibility, the solid line is a fit to the model as described in the text.}
  \label{dcsus}
\end{figure}

The DC magnetic susceptibility $\chi$(\textit{T})=\textit{M}/\textit{H} for a single crystal sample of CuDCl is shown in Fig.~\ref{dcsus}. The inset shows the low-temperature part of $\chi$(\textit{T}). We observe a broad maximum and a rapid decrease at lower temperatures, which indicates short range antiferromagnetic correlations for low-dimensional systems. The behavior is totally consistent with what is expected for a Heisenberg S=1/2 AF chain with a Bonner-Fisher maximum at $T_{max}$=7.5 K. There is no evidence of 3D long range order down to the temperature \emph{T}=1.8 K from the susceptibility data.

While the susceptibility data is well described by the Bonner-Fisher curve, such measurement can not distinguish which direction the chain axis follows. In view of the crystal structure, the chain axis is expected to be along the crystallographic \textbf{c} axis . To confirm that, we measured the magnetic dispersion relations both along the chain direction and perpendicular to the chain direction in the $(h,0,l)$ plane.

\begin{figure}
 \includegraphics[width=7.5cm,bbllx=110,bblly=115,bburx=550,
  bbury=690,angle=0,clip=]{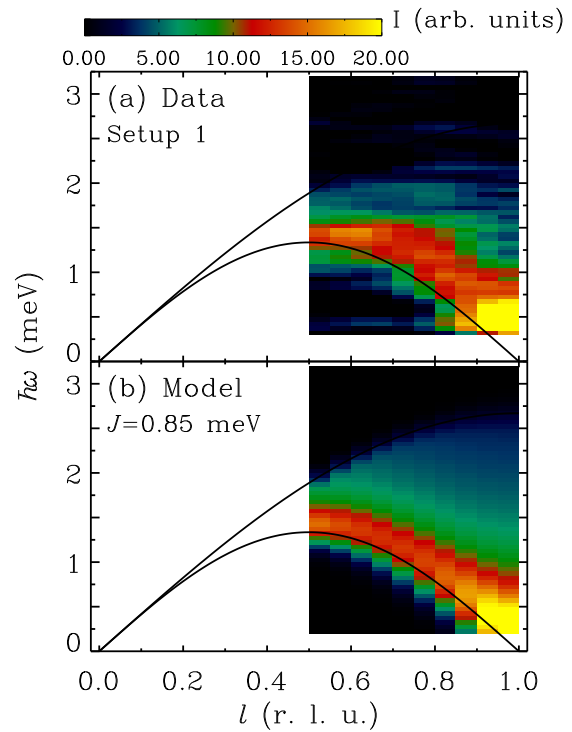}
 \caption{(color online) (a) False-color contour map of measured inelastic neutron scattering intensity in CuDCl as a function of transferred energy $\hbar\omega$ and transferred wavevector along the (l,0,\emph{l}) direction at \emph{T}=1.6 K, where the background is subtracted as described in the text. (b) Instrument resolution convolved model calculation of two-spinon continuum M$\ddot{u}$ller approximation with \emph{J}=0.85 meV . Solid lines are the predicted lower and upper bounds of the spinon continuum with \emph{J}=0.85 meV.}
  \label{contour}
\end{figure}

Fig.~\ref{contour}(a) shows the false-color map of background subtracted magnetic scattering intensity for two-spinon continuum in CuDCl, at \emph{T}=1.6 K using Setup 1. This data set was obtained by combining eleven constant-\textbf{q} (1,0,$l$) scans along the chain axis for 0.5$\leq l\leq$1 with step size 0.05. The background is determined as gaussian fits to the raw data which are at least 0.1 meV outside of the known bounds of the two-spinon continuum for CuDCl, which accounts for the incoherent elastic scattering of the sample and the thermal diffuse scattering of the analyzer.

\begin{figure}
 \includegraphics[width=7.5cm,bbllx=70,bblly=115,bburx=480,
  bbury=735,angle=0,clip=]{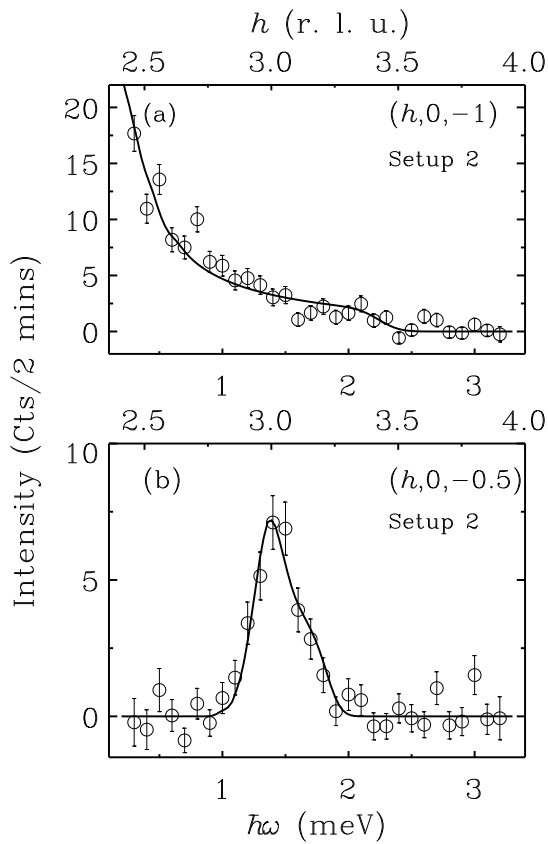}
 \caption{Typical scans of background-subtracted neutron scattering intensity vs transferred energy $\hbar\omega$ in CuDCl at \emph{T}=1.9 K (a) zone center (\emph{h},0,-l) and (b) zone boundary (\emph{h},0,-0.5). Solid lines are the fits to the data based on M$\ddot{u}$ller approximation of two-spinon continuum convolved with the instrument resolution.}
  \label{e-scan-L}
\end{figure}

INS measurements along the chain direction were also performed using Setup 2 with better energy resolution. Fig.~\ref{e-scan-L} shows the typical background-subtracted neutron scattering intensity versus $\hbar\omega$. Here the data were collected by using the focusing analyzer mode, where the value of \emph{h} was chosen so that the scattered wave-vector $\textbf{k}_\textbf{f}$ is parallel to the chain axis \textbf{c} to achieve the good wave-vector resolution along the chain. Fig.~\ref{e-scan-L}(a) at zone center \textbf{q}=(\emph{h},0,-1) shows the scattering intensity consistently decreasing, which indicates the divergence at the lower bound in the continuum. Fig.~\ref{e-scan-L}(b) at zone boundary \textbf{q}=(\emph{h},0,-0.5) shows a prominent peak at $\sim$1.4 meV, which is corresponding to the value of $\frac{\pi J}{2}$ and \emph{J} is the nearest neighbor intra-chain exchange constant.

\begin{figure}
 \includegraphics[width=7.5cm,bbllx=40,bblly=110,bburx=500,
  bbury=640,angle=0,clip=]{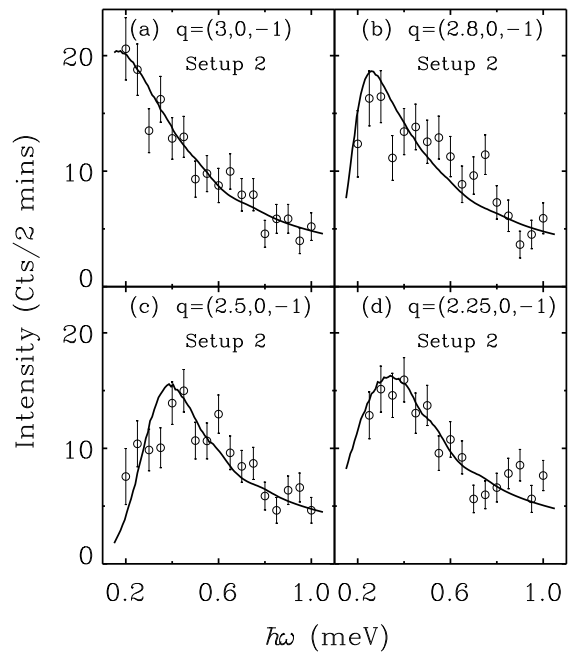}
 \caption{Typical scans of background-subtracted neutron scattering intensity vs $\hbar\omega$ in CuDCl at \emph{T}=1.9 K showing the weak dispersion of magnetic excitation with \emph{h}. Solid lines are the fits to the data as described in the text.}
  \label{e-scan-h}
\end{figure}

Furthermore, we measured the dispersion perpendicular to the chain axis using Setup 2. Representative background-subtracted energy scans up to 1 meV with wave-vectors perpendicular to the chain axis \textbf{c} are shown in Fig.~\ref{e-scan-h}. To determine the background, constant-\textbf{q} scans of (\emph{h},0,-0.5) and (\emph{h},0,-1.5) at the same configurations were measured for each \emph{h}, where no magnetic excitation is expected in this energy transfer range ($\hbar\omega\leq$1 meV) and the background is taken account as a fit to Gaussian profile. The apparent dispersion along \emph{h} suggests the existence of next-nearest-neighbor exchange constant $J'$ along the \textbf{a} axis.

The solid line in Fig.~\ref{dcsus} is a fit of measured DC susceptibility data to
\begin{eqnarray}
\chi=\chi_{dia}+f\times\chi_{imp}+(1-f)\times\chi_m,
\end{eqnarray}
where $\chi_{dia}$ is the temperature-independent diamagnetic contribution of the sample, $\chi_{imp}$ is the contribution from the impurity, $\chi_{m}$ is the spin susceptibility for 1D S=1/2 AFM chain and \emph{f} is the fraction of free spin due to the impurity. $\chi_{imp}$ simply obeys the Curie law. For $\chi_m$, a polynomial expression\cite{Estes80} which can reproduce the results of Bonner and Fish\cite{Bonner135:64} very well was used.

The fitting procedure yields $\chi_{dia}$=-2.4(1)$\times10^{-4}$ emu/Cu mol, \emph{f}=0.55(2)\%, and $J$=0.94(3) meV.

Since the numerical difference for the dynamic spin-correlation function ${\cal S}(\textbf{q},\omega)$ between the exact two-spinon solution\cite{Boug54:96,Karbach55:97} and M$\ddot{u}$ller approximation\cite{Muller24:81} is small, the data in Fig.~\ref{contour}(a) were globally fitted to the M$\ddot{u}$ller approximation of the 1D HAF S=1/2 chain convolved with instrumental resolution:
\begin{eqnarray}\label{sqw}
{\cal S}(\textbf{q},\omega)=\frac{1}{2\pi}\frac{A}{\sqrt{(\hbar\omega)^2-\epsilon_1^2(\textbf{q})}}
\\ \nonumber
\times\Theta(\hbar\omega-\epsilon_1(\textbf{q}))\Theta(\epsilon_2(\textbf{q})-\hbar\omega),
\end{eqnarray}
where A is an overall factor, $\Theta(x)$ is the step function, $\epsilon_1(\textbf{q})$=$\frac{\pi J}{2}|\sin(\pi l)|$ and $\epsilon_2(\textbf{q})$=$\pi J|\sin\frac{\pi l}{2}|$ are the lower and upper bounds of the two-spinon continuum. Fig.~\ref{contour}(b) shows the best fit of model simulation with \emph{J}=0.85(4) meV. The intensity is high at the lower bound due to the inverse square root divergence of the two-spinon density of states at that boundary.
The solid lines are the lower and upper two-spinon continuum boundaries with \emph{J}=0.85 meV.

To estimate the inter-chain coupling $J'$, we need to consider the case of weakly coupled AFM S=1/2 chains. Although the spin dynamics was well studied in the \textbf{\emph{ordered phase}} by using the mean-field (MF) approximation and random-phase-approximation (RPA)\cite{Schulz77:96,Essler56:97}, unfortunately so far there is no such theory to predict the spin dynamics including the transverse dispersion \emph{\textbf{above}} $T_N$. Nevertheless, we naively treat the transverse dispersion as a simple sinusoidal function. The lower bound of the continuum is modified to also include the dispersion along the transverse direction in the same way as the transverse dispersion term is included in spin wave below $T_N$:

\begin{eqnarray}
\epsilon_1^2(\textbf{q})=\frac{\pi^2 J^2}{4}\sin^2(\pi l)+\Delta^2\sin^2(\pi h),
\end{eqnarray}
where $\Delta$ is the bandwidth of the transverse dispersion. The expression of upper bound of the continuum is unchanged. The solid lines in Fig.~\ref{e-scan-L} and Fig.~\ref{e-scan-h} are the fits to the instrument convolution of the model with fixed \emph{J}=0.85 meV and the only adjustable parameter is the overall factor A. It agrees very well with the data and the fitting procedure gives $\Delta$=0.29(1) meV. We further assume that the bandwidth $\Delta$ is related to $J'$ as the same way in the transverse dispersion below $T_N$\cite{Schulz77:96}, $J'$ can be roughly estimated to be $\sim$$\Delta/6$=0.05 meV. Using the values of $J$ and $J'$, the Eq.~(13) in the Ref.[26] gives the estimation of $T_N\sim$1.3 K.

In summary, we have carried out x-ray crystallography, magnetic susceptibility and inelastic neutron scattering measurements to understand what is the right spin Hamiltonian of single-crystal samples in a newly synthesized compound CuDCl. In view of crystal structure, it consists of 1D S=1/2 HAF chains with the chain axis along the \textbf{c} axis. Bulk measurement and inelastic neutron scattering data are also consistent with the weakly coupled chain model calculation. Our future work will focus on the mixed magnetic system 2Dioxane$\cdot$2H$_2$O$\cdot$Cu(Cl$_{1-x}$Br$_x$)$_2$, which could be an ideal realization of S=1/2 chain with bond randomness that to date has only been theoretically studied.\cite{Doty45:92,Fisher50:94}

\begin{acknowledgments}
One of the authors T. Hong thanks the helpful discussion with M. B. Stone. Part of this work has been performed at the Paul Scherrer Institut. Research at ORNL was funded by the United States Department of Energy, Office of Basic Energy Sciences- Materials Science, under
Contract No. DE-AC05-00OR22725 with UT-Battelle, LLC. The work at
NIST is supported by the National Science Foundation under
Agreement Nos. DMR-9986442, -0086210, and -0454672.
\end{acknowledgments}

\end{document}